\documentclass[aps,prl,twocolumn,amsmath,amssymb,superscriptaddress]{revtex4-1}
\usepackage[dvips]{graphicx}% Include figure files
\usepackage{dcolumn}% Align table columns on decimal point
\usepackage{bm}% bold math
\usepackage{float,color}
\usepackage{gensymb}
\usepackage{hyperref}

\begin{document}

\title 
{
Superconducting single-layer $T$-graphene and novel synthesis routes
}

\author{Qinyan Gu}
\affiliation{
National Laboratory of Solid State Microstructures,
School of Physics and Collaborative Innovation Center of Advanced Microstructures,
Nanjing University, Nanjing, 210093, P. R. China}

\author{Dingyu Xing}
\affiliation{
National Laboratory of Solid State Microstructures,
School of Physics and Collaborative Innovation Center of Advanced Microstructures,
Nanjing University, Nanjing, 210093, P. R. China}

\author{Jian Sun}
\email[Correspondence should be addressed to J.S. ]{(E-mail: jiansun@nju.edu.cn)}
\affiliation{
National Laboratory of Solid State Microstructures,
School of Physics and Collaborative Innovation Center of Advanced Microstructures,
Nanjing University, Nanjing, 210093, P. R. China}

\date{\today}

\begin{abstract}

Single-layer superconductors are ideal materials for
fabricating superconducting nano devices. 
However, up to date,
very few single-layer elemental superconductors have been predicted and 
especially no one has been successfully synthesized yet.
Here, using crystal structure search techniques and {\it ab initio} calculations,
we predict that a single-layer planar carbon sheet with 4- and 8-membered rings 
called $T$-graphene is a new intrinsic elemental superconductor with superconducting critical 
temperature ($T_c$) up to around 20.8 K. 
More importantly, we propose a synthesis route to obtain such a single-layer $T$-graphene,
that is, a $T$-graphene potassium intercalation compound
(C$_4$K with $P4/mmm$ symmetry) is firstly synthesized at high pressure ($>$ 11.5 GPa) 
and then quenched to ambient condition;
and finally, the single-layer $T$-graphene can be either exfoliated using the electrochemical method
from the bulk C$_4$K, or peeled off from bulk $T$-graphite C$_4$, 
where C$_4$ can be obtained from C$_4$K by evaporating the K atoms.
Interestingly, we find that the calculated $T_c$ of C$_4$K 
is about 30.4 K at 0 GPa, which sets a new record for layered 
carbon-based superconductors.
The present findings add a new class of carbon-based superconductors. 
In particular, once the single-layer $T$-graphene is synthesized,
it can pave the way for fabricating superconducting devices 
together with other 2D materials using the layer-by-layer growth techniques.

\end{abstract}

\maketitle

%\section{Introduction}

Superconductivity in single layer materials \cite{Saito2016}, 
such as FeSe \cite{Wang2012-sinle-layer-FeSe-SC-CPL}, 
MoS$_2$ \cite{Lu2015}, NbSe$_2$ \cite{Xi2015}  
has attracted tremendous attention recently due to their inspiration in fundamental science
and potentials in future applications. 
However, without charge doping or other adjustments, 
very few examples of intrinsic single layer superconductor
have been found.
For instance, graphene is not a superconductor due to the vanishing density of states
at the Dirac point, although it can become a 2D single layer superconductor 
with charge doping or tensile strain \cite{Si2013, Ludbrook2015}.
Recently reported superconductivity in magic-angle graphene \cite{Cao2018}
caused a sensation, 
but a magic angle between two adjacent graphene layers is required, 
which is difficult to be controlled.
From what has been proposed so far, the 2D boronphene
seems to be the only intrinsic single layer superconductor
without external strain or charge doping. \cite{Penev2016, Zhao2016}
However, the synthesis of such a kind of 2D elemental superconductor remains a challenge.

Carbon is one of the most versatile elements 
and can constitute various types of molecules and crystals
due to its rich electronic hybridization configurations: sp, sp$^2$ and sp$^3$. 
In addition to the well-known diamond, graphite,
C$_{60}$ fullerene \cite{Kroto1985}, carbon nanotube \cite{Iijima1991} and graphene \cite{Novoselov2005},
many other types of carbon allotropes
have been studied,
including M-carbon \cite{Li2009-M-carbon}, bct-carbon \cite{Zhou2010}, 
graphdiyne \cite{Li2014}, nanotwin diamond \cite{Huang2014-nt-diamond-Nature}, 
penta-graphene \cite{Zhang2015},
V-carbon \cite{Yang2017-C70-peapods-PRL}, haeckelites \cite{Terrones2000}, 
etc, as an incomplete list.
Recently, a buckled $T$-graphene was predicted to have
Dirac-like fermions and a high Fermi velocity similar to graphene \cite{Liu2012}.
This new form of carbon sheet was predicted to 
present good mechanical properties \cite{Majidi2017}
and could be used in hydrogen storage \cite{Sheng2012}.

Although there is no superconductivity in pure graphite,
many carbon-allotrope related materials were reported to be superconducting, 
such as graphite intercalation compounds (GICs) \cite{Hannay1965}, 
fullerene alkali metal compounds \cite{Ganin2010}, 
nanotubes \cite{Tang2001}, boron-doped diamond \cite{Ekimov2004}, 
boron carbide \cite{Xia2017-B6C}
and magic-angle graphene superlattices \cite{Cao2018}.
Due to the similarity of the graphene-like boron layers 
in MgB$_2$ \cite{Nagamatsu2001}, 
GICs draw extensive attention, in which metallic atoms 
intercalate between the graphene sheets.
Alkali metal carbon compounds (C$_8$A, A=K, Rb, Cs)
were the first type of GIC superconductors \cite{Hannay1965} studied,
with $T_c$ being usually less than 1 K.
Since C$_8$K is one of the easiest to fabricate in alkali carbon compounds\cite{Hannay1965}, 
intense efforts have been taken to the study on
its superconducting properties \cite{Grueneis2009, Pan2011}.
With pressure, the $T_c$ of C$_8$K increases to 1.7 K at 1.5 GPa \cite{Smith2006}.
Up to now, among the GICs theoretically proposed, 
C$_6$Yb (with $T_c$ = 6.5 K) \cite{Weller2005} 
and C$_6$Ca (with $T_c$ = 11.5 K)  \cite{Emery2005}  
have the highest $T_c$ at ambient pressure. 
And C$_6$Ca exhibits $T_c$ = 15.1 K 
at high pressure of 7.5 GPa \cite{Gauzzi2007}.
In addition to the bulk GICs, 
superconductivity in doped graphene with
lithium\cite{Profeta2012-Li-graphene}, potassium \cite{ChenGF2012-K-graphene}
and Calcium \cite{Chapman2016}, etc has been studied theoretically and experimentally.
On the other hand, 
pressure is widely applied to explore new materials with unexpected stoichiometries
as well as structures. \cite{Zhang2013, Zhang2017-nature-review}
Considering the abundant carbon allotropes under normal or pressurized conditions,
one would ask the questions: Is there any new 2D carbon intercalation compound 
with better superconducting properties?
Can we find a 2D pure carbon allotrope
with intrinsic $T_c$?

In this Letter, using crystal structure search methods and first-principles 
calculations, we predict a very interesting C$_4$K compound with $P4/mmm$ symmetry.
It has 2D carbon layers with the structure as $T$-graphene, 
where potassium atoms intercalate in between, on the analogy of GICs.
We predict that this $P4/mmm$ C$_4$K can be synthesized above 11.5 GPa 
and is quenchable to ambient pressure.
exhibiting a high $T_c$ up to around 30.4 K at 0 GPa,
which is more than one order of magnitude higher than that of C$_8$K.
We also propose that $T$-graphene can be peeled off
from the bulk C$_4$K, once it is synthesized by high pressure method.
Most surprisingly, $T$-graphene itself is found to be an intrinsic 
2D elemental superconductor with $T_c$ of about 20.8 K at ambient pressure.

\begin{figure}[tph]
\begin{center}
\includegraphics[width=0.45\textwidth]{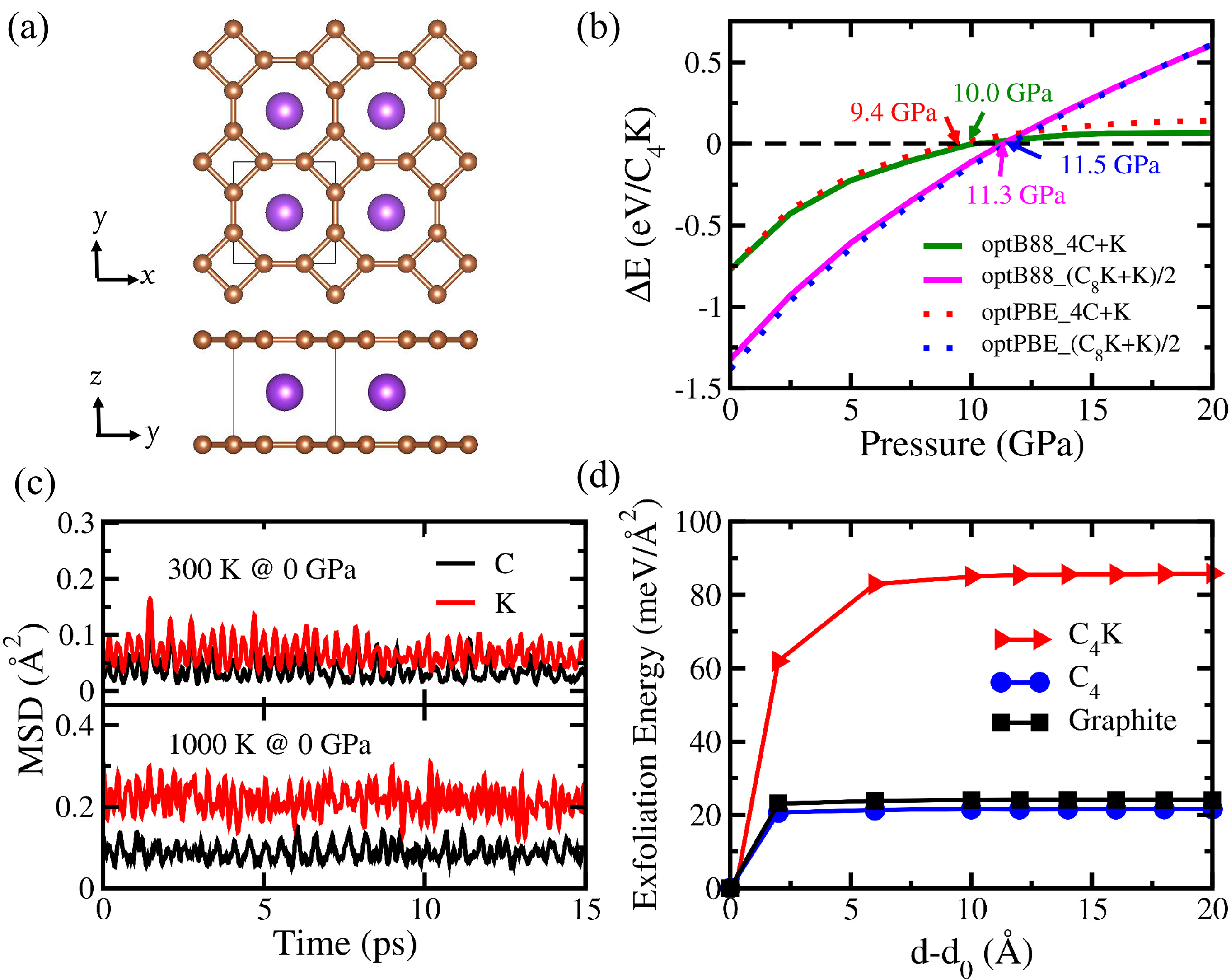}
\caption{
The crystal structure and stability of C$_4$K.
(a) The crystal structures of C$_4$K,
orange and purple balls denoting the C and K atoms, respectively.
(b) Calculated formation enthalpies for C + K or C$_8$K + K
relative to C$_4$K under high pressure,
with optB88-vdw \cite{Klimes2009} and optPBE-vdw \cite{Klimes2011} functionals.
(c) The mean square displacement (MSD) of the K and C atoms
from the AIMD simulations for C$_4$K
at ambient pressure and temperatures of 300 (upper panel) and 1000 K (lower panel).
(d) Calculated exfoliation energies of single-layer $T$-graphene from
C$_4$K and bulk $T$-graphite (C$_4$) versus
separation distance $d-d_0$, compared with that of graphene.
}
\label{fig:energetics}
\end{center}
\end{figure}

%\section{Results and discussions}
To find a possible route to synthesize $T$-graphene,
we first use our newly developed machine learning accelerated 
crystal structure search method \cite{Xia2018-WN6},
to explore possible stable phases of C-K system up to 20 GPa.
More details about the crystal structure search and 
{\it ab initio} calculations can be found in the method section 
and the Supplemental Material (SM).
We find that the known C$_8$K compound becomes unstable at 20 GPa 
and tends to decompose to more stable C$_4$K and other compounds. 
A C$_4$K compound with $P4/mmm$ structure becomes 
energetically favorable at this pressure, in which 
carbon atoms construct a 2D sheet with 4- and 8-membered rings,
named as $T$-graphene.
While the K atoms in C$_4$K occupy the interlayer site above the centers of the 
octagonal ring (shown in Fig.\ \ref{fig:energetics}(a)). 
Due to the unique layered structure of C$_4$K and
its similarity to GICs, 
we believe it belongs to a new type of intercalation compounds, 
and we call them the $T$-graphene intercalation compounds (TGICs).

To further confirm the stability of $P4/mmm$ C$_4$K, 
we calculate its formation enthalpy 
in two different possible synthesis routes,
$4C + K \rightarrow C_{4}K$ and $C_{8}K + K \rightarrow 2C_{4}K$, 
ranging from 0 to 20 GPa.
Under different pressures,
the most stable phases of the elements (graphite or diamond for carbon,
BCC or FCC for potassium) are taken 
as the references to calculate the formation enthalpy.
A hard version pseudopotential with small core radium for carbon and
a potassium pseudopotential with 9 valence electrons 
(including all the 3$s$, 3$p$ and 4$s$ electrons), 
together with an extremely high cutoff energy (1050 eV)
and very dense k-mesh are used. 
More details can be found from the method section and the SM.
As shown in Table S1 in the SM,
the optB88-vdw \cite{Klimes2009} and optPBE-vdw \cite{Klimes2011} functionals
yield the best agreement with the experimental lattice constants for graphite
and potassium, as well as the correct ground state of potassium. 
Therefore, we use them to calculate the formation enthalpy of C$_4$K.
As shown in Fig.\ \ref{fig:energetics}(b), 
although the transition pressures from different functionals are slightly different, 
the conclusions are consistent that
C$_4$K is more stable relative to both $C + K$ and $C_{8}K + K$ under high pressure.
The highest transition pressure is around 11.5 GPa.
We calculate the phonon spectra of $P4/mmm$ C$_4$K 
at 0 GPa and 20 GPa, see Fig. S3 in the SM, 
and find that both of them are stable. 
This indicates that $P4/mmm$ C$_4$K is possible to be synthesized 
at high pressure and quenchable to ambient pressure.
To check the stability of C$_4$K at finite temperature, 
we perform {\it ab initio} molecular dynamics simulations.
As shown by the mean square displacement (MSD) results in Fig.\ \ref{fig:energetics}(c), 
C$_4$K is also stable at ambient pressure and at finite temperature (300 K and 1000 K).
Nevertheless, the synthesis of C$_4$K may still be challenging and it requires
cutting-edge techniques that provide high pressure and high temperature condition, 
such as laser-heated diamond anvil cells \cite{Salamat2014-LHDAC-review}
or large-volume multianvil apparatus.
These techniques have been used to successfully synthesize
many new materials, including carbon-based materials
\cite{Huang2014-nt-diamond-Nature, Yang2017-C70-peapods-PRL, Salamat2014-LHDAC-review}.

As shown by convex hulls in Fig. S1(a) in the SM,
we find that there are three new stable/metastable 
candidates of C-K compounds at 20 GPa:
the C$_4$K with $P4/mmm$ structure, the C$_3$K$_4$ with $C2/m$ structure,
and the C$_2$K$_3$ with $C2/m$ structure.
The C$_8$K becomes energetically unfavorable and tends to decompose to C$_4$K and other compounds
when pressure increases to 20 GPa.
We note that the carbon atoms in C$_3$K$_4$ and C$_2$K$_3$
form chain-like structures (see Fig. S2 in the SM).

As we know, liquid exfoliation and mechanical cleavage haven been
widely used to
get 2D monolayers from their laminated bulk crystals
with weakly coupled layers \cite{Novoselov2004, Liu2014}.
Exfoliation energy is the energy cost which is used to
peel off the topmost single layer from the surface of bulk crystal.
To explore the feasibility of peeling,
we calculate the exfoliation energy of $T$-graphene 
from bulk $P4/mmm$ C$_4$K and $T$-graphite C$_4$, respectively,
where bulk $T$-graphite C$_4$ can be obtained from C$_4$K by evaporating the potassium atoms,
possibly using similar procedure in the synthesis of Si$_{24}$ from Na$_{4}$Si$_{24}$\cite{Kim2015}.
As shown in the Fig. S5 in the SM,
the potential energy barrier of moving the K atom in C$_4$K 
is very similar to that of the Na atom in Na$_{4}$Si$_{24}$,
which shows the feasibility of obtaining bulk C$_4$ from C$_4$K.
We take the graphene peeled off from bulk graphite as a reference system
and the exfoliation energy of graphene
is calculated to be around 24.1 meV/$\AA^2$,
which is in good agreement with the experimental results 
(20 $\pm$ 1.88 meV/$\AA^2$) \cite{Zacharia2004, Ziambaras2007}.
Here, we use a vacuum layer of 20 {\AA}
to make sure that the charge density of the two surfaces do not overlap.
We then used the same parameters to calculate the exfoliation energy of $T$-graphene.
With different parent materials of bulk $P4/mmm$ C$_4$K and $T$-graphite C$_4$,
the resulted exfoliation energy are  
85.8 meV/$\AA^2$ and 21.6 meV/$\AA^2$, respectively,
as shown in Fig.\ \ref{fig:energetics}(d).

\begin{figure}[tph]
\begin{center}
\includegraphics[width=0.45\textwidth]{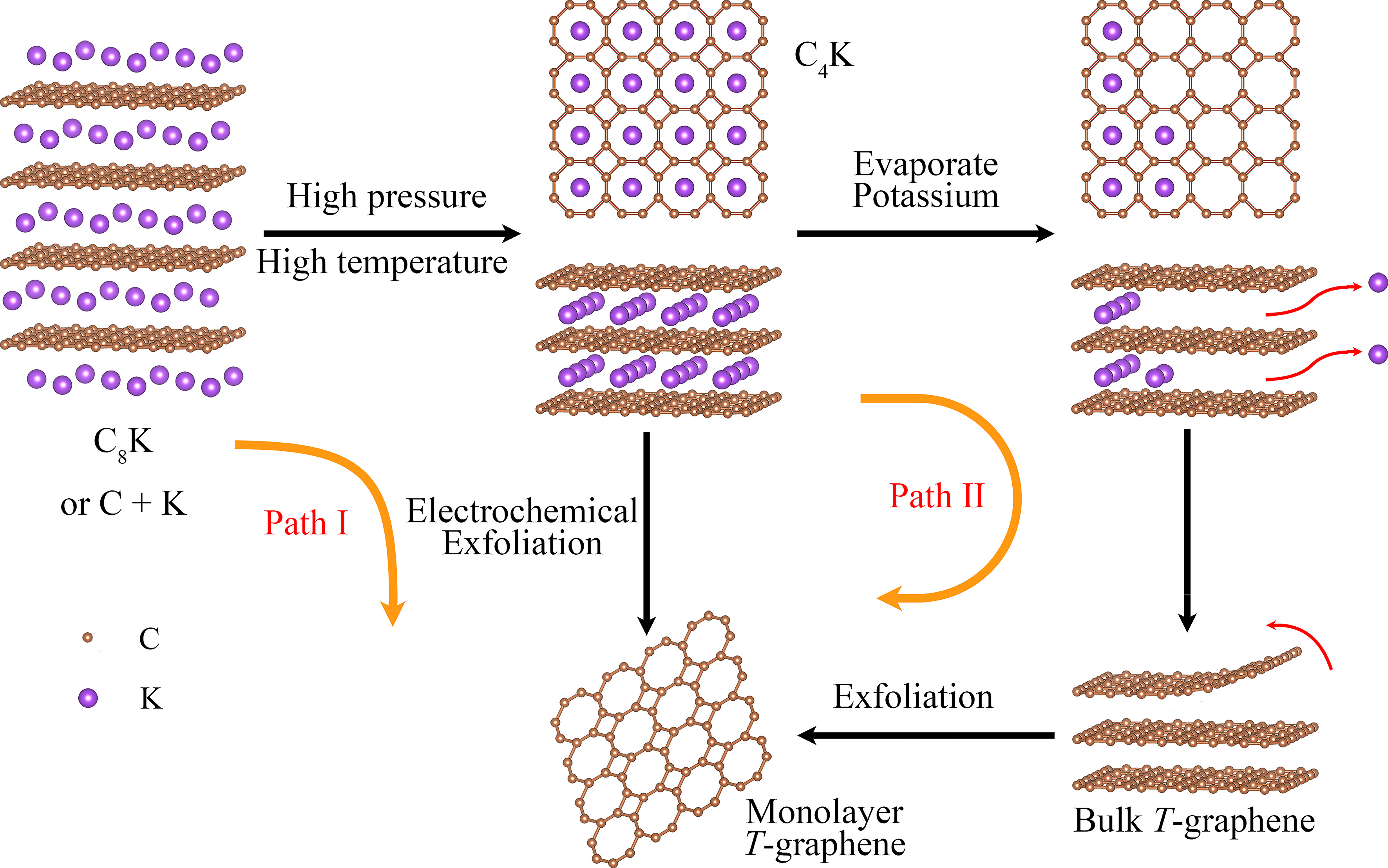}
\caption{
A sketch map of the proposed synthesis routes for the superconducting single-layer $T$-graphene.
Path I represents the electrochemical exfoliation route from C$_4$K and
Path II represents the mechanical exfoliation route from the bulk $T$-graphite,
where the bulk $T$-graphene (C$_4$) can be obtained from C$_4$K by evaporating the potassium atoms
with a similar method used in the synthesis of Si$_{24}$ from Na$_{4}$Si$_{24}$\cite{Kim2015}.
}
\label{fig:synthesis-routes}
\end{center}
\end{figure}

According to the literature \cite{Mounet2018},
a 2D material is potentially exfoliable (or easily to be exfoliated)
when the exfoliation energy is $<$ 130 meV/$\AA^2$
($<$ 30-35 meV/$\AA^2$). 
Therefore, $T$-graphene is potentially exfoliable from C$_4$K and 
much easier to be exfoliated from bulk $T$-graphite C$_4$.
As shown in Table.\ \ref{table:charge-transfer},
we find that C$_4$K has similar charge transfer as 
that in many other graphite intercalation compounds, 
such as C$_8$Li, C$_8$Na, C$_8$K and C$_6$Li.
Although having ionic-like interactions, 
it was reported that graphene sheets can be exfoliated from C$_6$Li, C$_8$Na and C$_8$K
in experiments using the electrochemical exfoliation method.
\cite{Lee2015, Bharath2017, Thomas2018}
Thus due to the structural similarity and also the similar charge transfer,
$T$-graphene can very likely be exfoliated from bulk C$_4$K 
with similar electrochemical method 
as used in the graphite intercalation compounds mentioned above.
The routes we proposed to synthesize 
monolayer $T$-graphite are summarized in Fig.\ \ref{fig:synthesis-routes}.

\begin{table}[bp]
\centering
  \caption{
The amount of charge transfer from the intercalated metal atoms to the carbon sheets 
in C$_4$K and some typical graphite intercalation compounds.
}
\begin{tabular}[t]{m{1.5cm}m{1.3cm}m{1.3cm}m{1.3cm}m{1.3cm}m{1.3cm}}
        \hline\hline
         System & C$_4$K & C$_6$Li & C$_8$Li  & C$_8$Na  & C$_8$K \\
        \hline
        Charge transfer & 0.8$e$  & 0.9$e$ & 0.8$e$ & 0.9$e$ & 0.8$e$ \\
      \hline\hline
   \end{tabular}
\label{table:charge-transfer}
\end{table}

\begin{figure}[thp]
\begin{center}
\includegraphics[width=0.45\textwidth]{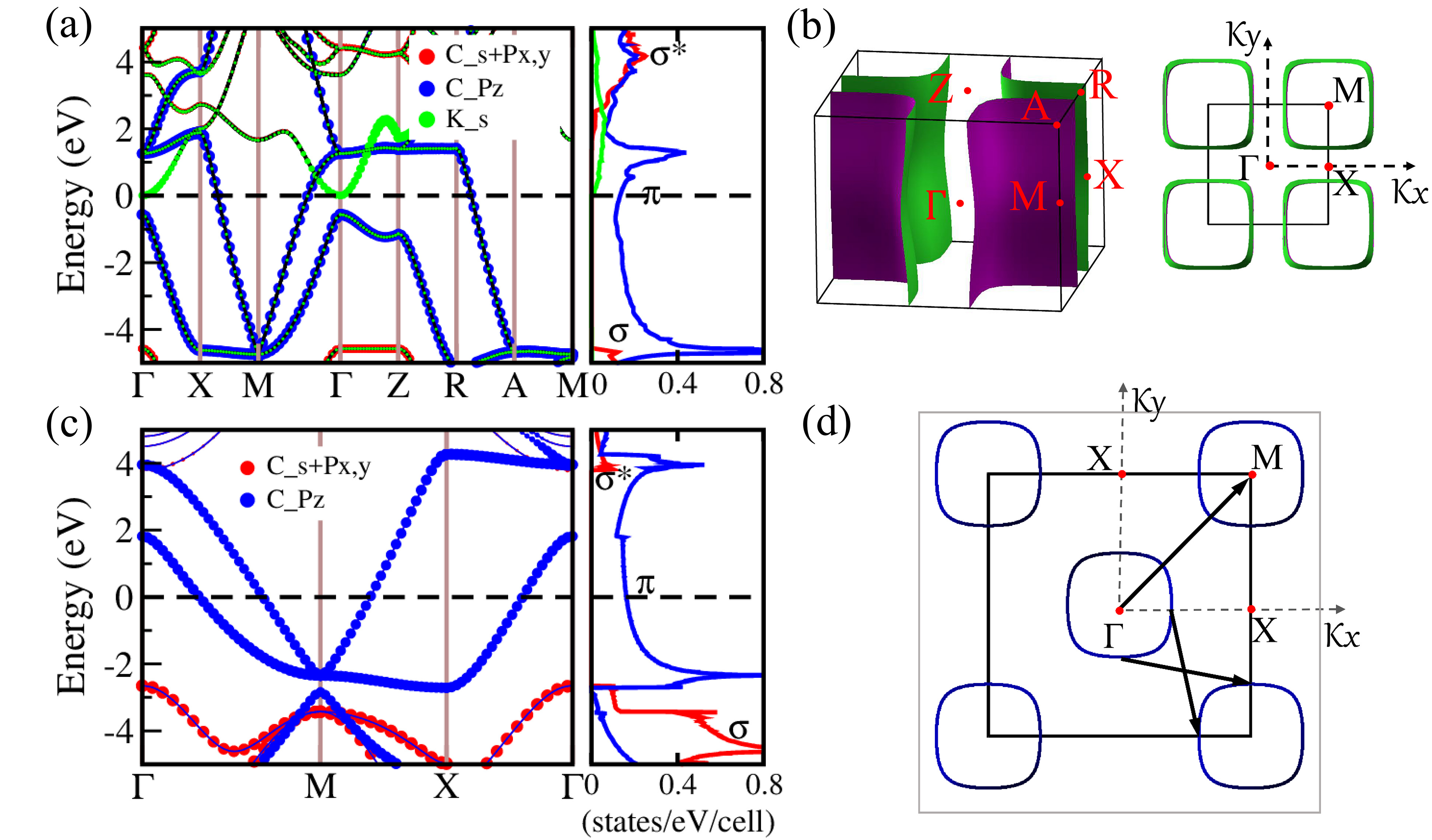}
\caption{
Calculated electronic structures of C$_4$K and single-layer $T$-graphene at 0 GPa.
The orbital-resolved band structures (left panel) and 
electron DOS (right panel) are shown in (a) for C$_4$K and (c) for $T$-graphene. 
The size of the dots in the band structures
stands for the weights of different orbitals. 
The Fermi level is denoted by the dashed horizontal line.
(b) Front view (left) and the top view (right) of the Fermi surface of C$_4$K
in the extended Brillouin zone. 
(d) Fermi contour of $T$-graphene in the extended BZ.
The bold rectangle represents the conventional BZ 
and the arrows point to the nesting vectors.
}
\label{fig:band}
\end{center}
\end{figure}

In what follows we investigate the electronic properties
of bulk C$_4$K and single-layer planar $T$-graphene.
The calculate electronic structures of C$_4$K and $T$-graphene 
are plotted in Fig.\ \ref{fig:band}.
We also calculated the band structures of the C$_4$ layer without potassium atoms (C$_4K_0$),
which is obtained by removing the K atoms from the C$_4$K structure and
keeping the lattice constants at 0 GPa 
(see Fig. S6 in the SM).
Compared with C$_4$K$_0$, 
the intercalation of K atoms pushes up the Fermi level 
and increases the occupancy of electrons in $\pi$ bands.
From Fig.\ \ref{fig:band}, one can see that either in C$_4$K or $T$-graphene, 
the electronic density of states at the Fermi level, 
$N\left ( \epsilon _{F} \right )$, is mostly dominated by the $\pi$ bands.
In C$_4$K, there is only one band crossing the Fermi level, 
which results in a relatively smooth cylinder-like Fermi surface
around the $M$ point, as it can be seen in Fig.\ \ref{fig:band}(b).
While in the $T$-graphene, the absence of K atoms lowers the Fermi level
and adds a new hole pocket around the $\Gamma$ point.
As shown in Fig.\ \ref{fig:band}(b) and (d),
both fermi surfaces of C$_4$K and $T$-graphene have 
2D features and perfect nesting vectors.

\begin{figure}[thp]
\begin{center}
\includegraphics[width=0.45\textwidth]{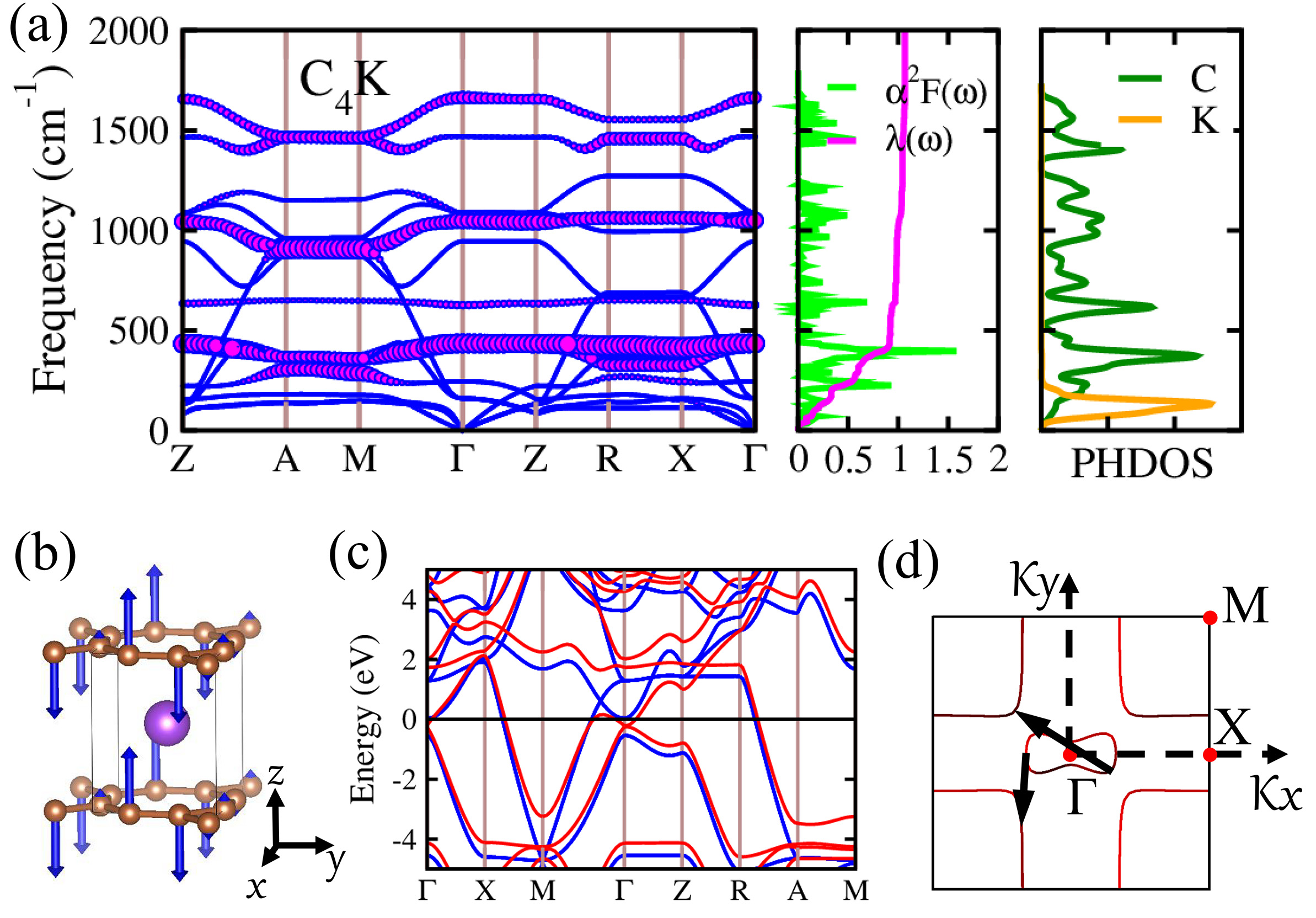}
\caption{
Electron-phonon coupling and electronic structures with/without perturbation of C$_4$K at 0 GPa.
(a) Phonon dispersion curves, Eliashberg spectral functions $\alpha^2F$($\omega$)
together with the electron-phonon integral $\lambda$($\omega$)
and phonon density of states (PHDOS).
(b) the E$_g$ vibrational mode.
(c) The electronic band structures of C$_4$K 
in the presence (red) and absence (blue) of the perturbation from the E$_g$ mode.
(d) Fermi surface in the $k_z = 0$ plane after perturbation
and the arrows point to the nesting vectors.
}
\label{fig:elph-C4K}
\end{center}
\end{figure}

Since the Fermi contours of C$_4$K and $T$-graphene have similar cylinder shapes 
as those in the cuprates \cite{Damascelli2003} 
and iron-based superconductors \cite{Kuroki2008}, 
it is natural to guess if they are also superconducting.
Therefore, we investigate their superconducting properties by
electron-phonon coupling calculations and 
the Allen-Dynes modified McMillian equation \cite{Allen1975}.
We estimate that the $T_c$ of $T$-graphene is around 20.8 K at ambient pressure
with a commonly used screened Coulomb potential $\mu^*$ = 0.1.
And in C$_4$K, its $T_c$ is around 30.4 K at 0 GPa, 
which is the highest among all known graphite intercalation compounds.
One can consider that the number of doping atoms in C$_4$K is more than that in C$_8$K,
thus it seems that a higher doping level is a positive factor
to the superconductivity in these carbon-sheet based systems.

In Fig.\ \ref{fig:elph-C4K}, we plot phonon dispersions together with phonon linewidth,
Eliashberg spectral function $\alpha^2F$($\omega$),
electron-phonon coupling strength $\lambda$($\omega$)
and phonon density of states (PHDOS) of C$_4$K at ambient pressure.
According to the electronic structures discussed above, 
there is only one electron pocket in the Fermi surface of C$_4$K,
which is very similar to the case of single-layer FeSe 
on SrTiO$_3$ \cite{He2013-SL-FeSe-ARPES-NM}.
According to $\lambda =2\int \frac{\alpha ^{2}F\left ( \omega  \right )}{\omega }d\omega $, 
there is a negative correlation between $\lambda$ and $\omega$, 
so that the low frequency vibrations usually can enhance
the electron-phonon coupling and thus the superconductivity.
In C$_4$K, most contributions to the electron-phonon coupling constant
arise from the low-frequency modes, especially those related to the vibration of the K atoms
and the out-of-plane $E_g$ mode of the carbon sheet,
as shown in Fig.\ \ref{fig:elph-C4K}(b).
This also corresponds to the two predominant peaks 
in the Eliashberg spectral functions $\alpha^2F$
at about 300 and 450 $cm^{-1}$.

To provide more insights into the electron-phonon coupling,
we also compared the electronic bands without and with
the perturbation induced by the $E_g$ phonon mode with $\Delta Q$ = 0.1 $\AA$.
The calculated band structures and Fermi surfaces with the perturbation
are plotted in Figis. \ref{fig:elph-C4K} (c) and (d), respectively.
Compared to the unperturbed case,
the perturbated interlayer band near $\Gamma$ point is dramatically lowered by 0.3 eV
leading to a decrease in electronic occupation of the $\pi$ bands.
Particularly, the unperturbated interlayer bands have a gap 
near the Fermi level around the $\Gamma$ point, 
while the perturbated bands get in touch together and produce
a new electron pocket at the $\Gamma$ point.
These results clearly show that 
the low-energy carbon out-of-plane vibrations 
are critical to the electron-phonon pairing,
similar to the case in GICs.
Under pressure, the reduced interlayer distance between 
the carbon sheet and K atoms will push up the bands of K atoms 
(see Fig. S9 in the SM)
and decrease the electron-phonon coupling.
This argument is consistent with the present calculated results, where T$_c$ of 
C$_4$K decreases under pressure (see Fig. S10 in the SM).

\begin{figure}[thp]
\begin{center}
\includegraphics[width=0.45\textwidth]{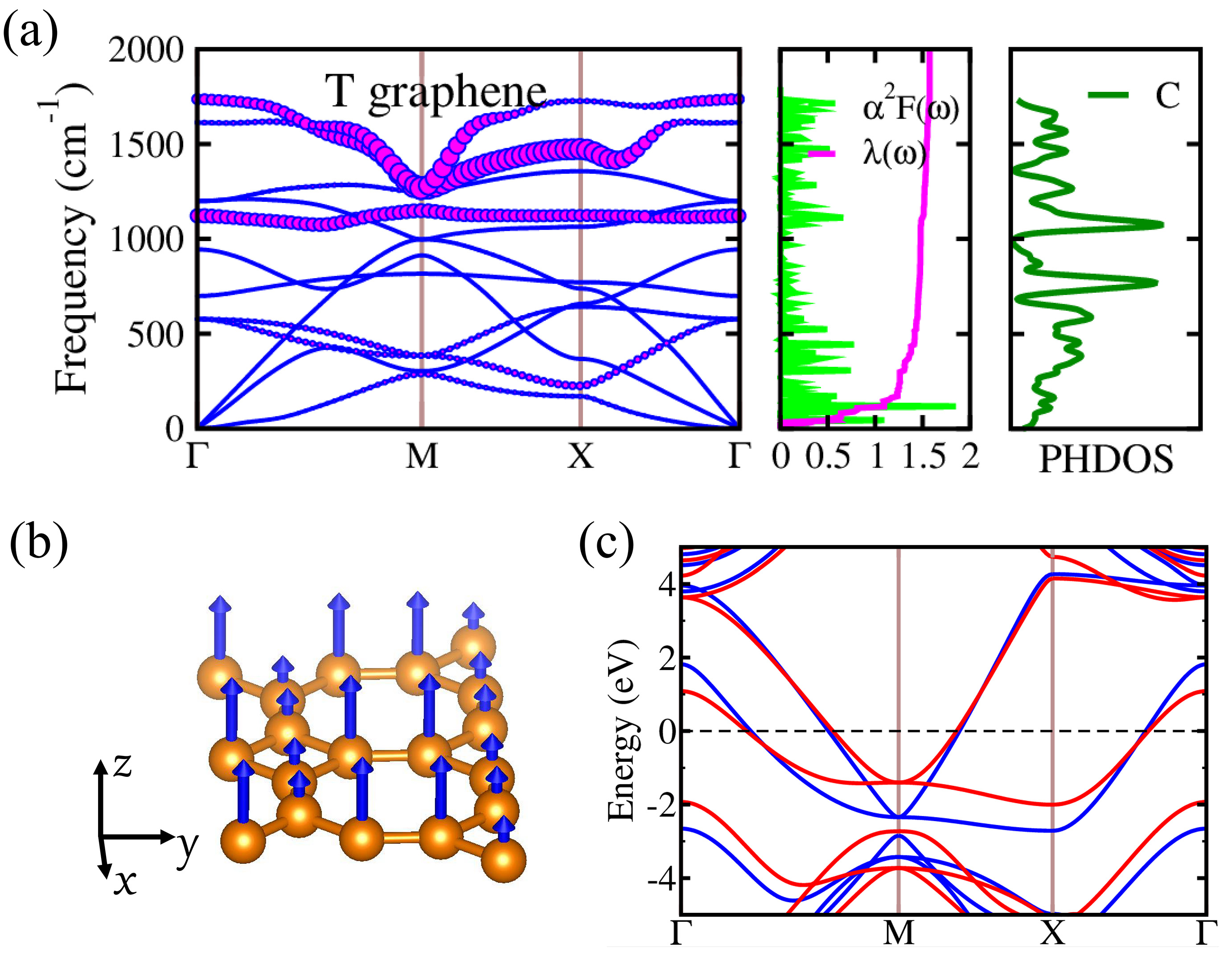}
\caption{
Electron-phonon coupling and electronic structures of 
single-layer $T$-graphene at 0 GPa.
(a) Phonon dispersion curves, Eliashberg spectral functions $\alpha^2F$($\omega$)
together with the electron-phonon integral $\lambda$($\omega$)
and phonon density of states (PHDOS).
(b) The vibrational A$_1$ mode.
(c) The electronic band structures in the presence (red) and absence (blue) of the perturbation from the A$_1$ mode.
}
\label{fig:elph-C4}
\end{center}
\end{figure}

Fig.\ \ref{fig:elph-C4} shows the electron-phonon coupling, 
electronic structures as well as the representative 
phonon mode of the single-layer T-graphene.
From Fig.\ \ref{fig:elph-C4}(a),
one can see that three acoustic modes 
in the low-frequency range ($\omega$ $\le$ 175 $cm^{-1}$)
couple strongly with electrons on the Fermi surface,
which makes a great contribution to $\lambda$ and yields $\lambda \approx$ 1.23, 
almost 78.3\% of the total electron-phonon coupling constant.
Among these three acoustic modes, the softest out-of-plane mode 
has the largest phonon linewidth,
which gives rise to the highest peak in the Eliashberg spectral functions.
We plot the vibration of the softest acoustic mode at the $X$ point 
($A_1$ mode with $\omega$ $\sim$ 160 $cm^{-1}$)
in Fig.\ \ref{fig:elph-C4}(b).
To reveal its effect on the electrons on the Fermi-surface,
in Fig.\ \ref{fig:elph-C4}(c),
we show the calculated electronic bands with and without the
perturbation induced by the $A_1$ phonon mode with $\Delta Q$ = 0.1 $\AA$.
With the perturbation, the band dispersions
and Fermi velocities near the Fermi level have obvious changes,
indicating the coupling between the electrons and this softest phonon mode.

In Fig.\ \ref{fig:elph-C4}(a),  
although the high frequency phonon modes above 1000 $cm^{-1}$
have a large linewidth,
they seem to have less contribution to the Eliashberg spectral functions
and also the total electron-phonon coupling constant $\lambda$ 
compared to the low frequency modes discussed above.
As shown in Fig. S7 in the SM,
the band dispersions in the presence of the perturbation induced by the $B_3u$ phonon at $X$ point 
are almost consistent with those in the absence of perturbation.
Therefore, the influence of the high-frequency carbon in-plane vibrations 
on the Fermi-surface electrons is weaker and they do not 
make a significant contribution to the superconducting pairing.
Although the out-of-plane modes in single-layer $T$-graphene 
have similar frequencies as those in bulk C$_4$K, 
there have not been too much electronic states to couple with.

%\section{CONCLUSIONS}

In summary, 
using {\it ab initio} calculations,
we have found that single-layer 
$T$-graphene with 4- and 8-membered rings is an intrinsic elemental
2D superconductor with a $T_c$ of around 20.8 K,
in which the low frequency out-of-plane vibrational acoustic modes
play a key role in superconducting pairing.
We have also proposed a novel route to synthesize the single-layer $T$-graphene, 
that is, first synthesize the $T$-graphene intercalation compounds 
by high pressure method,
and then exfoliate the single-layer $T$-graphene using electrochemical or other methods. 
As an example, we have searched carefully the C-K system 
using our machine learning accelerated crystal structure search method 
and find a $P4/mmm$ C$_4$K, which is exactly the $T$-graphene intercalation compound we want.
This C$_4$K compound can be synthesized when the pressure is higher than 11.5 GPa,
and can be quenched to ambient pressure.
Our calculation results show that the $T$-graphene should be feasible to be exfoliated from C$_4$K
using the electrochemical exfoliation method
once C$_4$K is synthesized by high pressure method.
Or be peeled off from bulk $T$-graphite C$_4$, 
where C$_4$ can be obtained from C$_4$K by evaporating the potassium atoms.
Interestingly, it is found that the calculated
$T_c$ of $P4/mmm$ C$_4$K is about 30.4 K at 0 GPa, 
which sets a new record for the layered carbon-based superconductors.

The coupling strength between the interlayer electronic states 
and carbon out-of-plane vibrations has the decisive effect on the superconducting properties 
of layered carbon intercalation compounds.
From the strong electron-phonon coupling, 
it follows that both $T$-graphene and C$_4$K exhibit conventional phonon-mediated superconductivity. 
Comparing the electronic structures and superconducting properties 
between $T$-graphene and C$_4$K, we can see the importance of doping effect. 
Therefore, a further enhancement on the $T_c$ for the single layer $T$-graphene
may be possible by charge doping, for instance, by doping metallic atoms or tuning gate voltage.
Defects may affect the superconducting properties of $T$-graphene and C$_4$K,
which is interesting to be investigated in future studies.
As one of the very few examples of intrinsic elemental single-layer superconductors, 
the $T$-graphene is an ideal 2D material that can be used to fabricate 
superconductor/semiconductor heterojunctions
with other 2D materials using the so-called ``vertical'' techniques.
This will greatly promote the development of the field.

\vspace{0.25cm}
{\it Method}
We use the machine learning accelerated
crystal structure search method \cite{Xia2018-WN6}
to investigate the stable structures of C-K system at 20 GPa,
with system sizes up to 18 atoms per simulation cell.
Structural optimizations, electronic band structure calculations
and strength calculations are
performed by the projector augmented wave (PAW) method
implemented in the Vienna {\it ab initio} simulation package (vasp) \cite{Kresse1996}.
In the structure searching, the generalized gradient approximation (GGA),
and the Perdew-Burke-Ernzerhof (PBE) functional \cite{Perdew1996}
are employed.
In the formation enthalpy calculations, a hard version pseudopotentials
with very small core radium for carbon and
a 9-valence-electron (including all the 3s, 3p and 4s electrons)
pseudopotential for potassium are used.
Together with an extremely high cutoff energy (1050 eV)
and very dense Monkhorst-Pack \cite{Monkhorst1976} k-sampling
using a small k-spacing of $0.02\times{2\pi}$\AA$^{-1}$.
Due to the 2D features of graphite, C$_4$K and other related intercalated compounds,
it is important to include the van der Waals (vdW) interaction to reproduce the
correct lattice constants and phase order.
We compare the results from different vdW corrections, 
including the Grimme's DFT-D2\cite{Grimme2006},
DFT-D3 \cite{Grimme2010},
optB88-vdW \cite{Klimes2009}, optPBE-vdW \cite{Klimes2011},
vdW-DF \cite{Dion2004, Roman-Perez2009}
and vdW-DF2 \cite{Lee2010},
with the experimental measured lattice constants of graphite and bcc potassium.
As listed in Table S1 in the SM, 
optB88-vdW and optPBE-vdW give relatively better agreement
and also give the correct ground state of potassium at ambient pressure,
which should be bcc but not fcc.
To check the stability of C$_4$K at finite temperature,
we perform {\it ab initio} molecular dynamics (AIMD) simulations
at ambient pressure and temperatures of 300 and 1000 K for 15 picoseconds
with a time step of 1 fs using the $NVT$ ensemble 
with Nose−Hoover thermostat \cite{Hoover1985}
within a supercell containing 135 atoms.
Electron-phonon coupling (EPC) calculations are performed
in the framework of Density functional perturbation theory (DFPT),
as implemented in the {\it quantum-espresso} code \cite{Giannozzi2009}. 
For C$_4$K , we adopt a $16\times16\times16$ k-point mesh
for the charge self-consistent calculation,
and a $32\times32\times32$ k-point mesh for EPC linewidth integration
and a $8\times8\times8$ q-point mesh for dynamical matrix.
For $T$-graphene, we adopt a $16\times16\times1$ k-point mesh
for the charge self-consistent calculation,
and a $32\times32\times2$ k-point mesh for EPC linewidth integration
and a $8\times8\times1$ q-point mesh for Dynamical matrix.
Norm-conserving Pseudopotentials are used with the energy cutoffs of 160 Ry
for the wave functions and 640 Ry
for the charge density to ensure that the converge criteria of total energy 
is less than 10$^{-6}$ Ry.

\vspace{0.25cm}
\noindent\textbf{Acknowledgments}\\
We thank for the fruitful discussions with Tong Chen,
Pengchao Lu, Kang Xia, Cong Liu, Yong Wang, Dexi Shao.
We appreciate the financial support from the MOST of China
(Grants No. 2016YFA0300404 and No. 2015CB921202),
the NSFC (Grant Nos. 11574133 and 11834006),
the NSF of Jiangsu Province (Grant No. BK20150012),
the Fundamental Research Funds for the Central Universities,
the Science Challenge Project (No. TZ2016001)
Calculations were performed on the
computing facilities in the High Performance Computing Center of
Collaborative Innovation Center of Advanced Microstructures,
the High Performance Computing Center of Nanjing University
and "Tianhe-2�at NSCC-Guangzhou.

%\bibliographystyle{apsrev}
%\bibliography{C4K}

\end{document}